\documentclass[a4paper,fleqn,10pt,twocolumn]{article}
\usepackage{hyperref}
\usepackage{amsmath}
\usepackage{amssymb}
\usepackage{array}
\usepackage{calc}
\usepackage{graphicx}
\usepackage{feynmp}
\usepackage{cite}
\newcommand{\mypreprint}[1]{\begin{flushright}
  {\large #1}\end{flushright}\vspace*{2.5ex}}
\newcommand{\mytitle}[1]{{\flushleft\huge\sffamily\bfseries #1}}
\newcommand{\myauthor}[1]{{\flushleft\normalsize #1}}
\newcommand{\myinstitute}[1]{\flushleft{\small #1}\\\vspace*{5ex}}
\newcommand{\myabstract}[1]{\centerline{
  {\large\sffamily\bfseries Abstract:\hspace*{1ex}}\parbox[t]{
    \textwidth*5/6-\widthof{{\large\sffamily\bfseries Abstract:}
    \hspace*{1ex}}}{#1}}\vspace*{5ex}}

\newcommand{\mysection}[1]{\section{\Large\sffamily\bfseries #1}}

\newcommand{\myssection}[1]{\section*{\Large\sffamily\bfseries #1}}

\newcommand{\rbr}[1]{\left( #1\right)}
\newcommand{\abr}[1]{\langle #1\rangle}
\newcommand{\cbr}[1]{\left\{ #1\right\}}
\newcommand{\sbr}[1]{\left[ #1\right]}
\newcommand{\done}{{\rm d}}

\newcommand{\dfour}{{\rm d}^4}

\newcommand{\tder}[2]{\frac{\done #1}{\done #2}}

\newcommand{\normalfeynmf}{
  \fmfset{thick}{1.25thin}
  \fmfset{arrow_len}{3mm}\fmfset{curly_len}{2mm}
  \fmfset{wiggly_len}{2mm}\fmfset{decor_size}{2mm}
  \fmfset{dot_size}{2thick}\unitlength=1.mm}

\newcommand{\myfigure}[3]{
  \begin{figure}[#1]
    \begin{center}
      #2\\\myfigcaption{\widthof{#2}}{#3}
    \end{center}
  \end{figure}
}

\begin{document}
\begin{fmffile}{bfkl_fg}
\newcommand{\angularorderingdis}[4]{
  \parbox{#3mm}{\begin{center}#4
      \begin{fmfgraph*}(#1,#2)
        \fmfstraight
        \fmftop{t1,t2,t3,t4,t5,t6}
        \fmfleft{l1,l2,l3,l4}
        \fmfright{r1,r2,r3,r4}
        \fmfbottom{b0,b1,b2,b3,b4,b5,b6,b7,b8}
        \fmf{fermion,tension=2,label.side=left,label.dist=.07h,
          label=$P$}{l2,v1}
        \fmf{fermion,tension=2,label.side=left,label.dist=.05h,
          label=$p_0$}{v1,vi1}
        \fmf{dots,tension=3}{vi1,vi2}
        \fmf{fermion,tension=2}{vi2,v11}
        \fmf{fermion,tension=2,label.side=left,label.dist=.05h,
          label=$p_{n-1}$}{v11,v12}
        \fmf{fermion,tension=1.5}{r2,v2}
        \fmf{phantom}{v2,v12}
        \fmffreeze
        \fmf{fermion,label.side=left,label.dist=.05h,
          label=$p_n$}{v12,v3}
        \fmf{photon,label.side=right,label.dist=.05h,
          label=$q$}{v2,v3}
        \fmf{fermion,tension=2}{v2,b6}
        \fmf{gluon,label.side=left,label.dist=.07h,
          label=$k_{n-1}$}{b4,v11}
        \fmf{gluon,label.side=left,label.dist=.07h,
          label=$k_n$}{b5,v12}
        \fmf{fermion}{v3,t4}
        \fmf{fermion}{v3,t5}
        \fmfv{decor.shape=circle,decor.size=.2h,decor.filled=0}{v1}
        \fmfv{decor.shape=circle,decor.size=.3h,decor.filled=0,
          label.dist=0,label=$\mu_F$}{v3}
        \fmffreeze
        \fmfi{plain}{vpath(__l2,__v1)shifted(thick*(1,2))}
        \fmfi{plain}{vpath(__l2,__v1)shifted(thick*(1,-2))}
      \end{fmfgraph*}
    \end{center}} 
}
\newcommand{\levs}[4]{
\parbox{#3mm}{\begin{center}#4
    \begin{fmfgraph*}(#1,#2)
      \fmfstraight
      \fmfbottom{l1,d1,la1,la,d2,ln}
      \fmftop{r1,d3,ra1,ra,d4,rn}
      \fmf{plain,label.side=left,label.dist=.05w,
        label=$p_a$}{l1,if1}
      \fmf{plain,label.side=left,label.dist=.05w,
        label=$k_1$}{if1,r1}
      \fmf{plain,label.side=right,label.dist=.05w,
        label=$p_b$}{ln,ifn}
      \fmf{plain,label.side=right,label.dist=.05w,
        label=$k_n$}{ifn,rn}
      \fmffreeze
      \fmf{dots}{if1,c1}
      \fmf{plain,label.side=right,label.dist=.045w,
        label=$q_{i-1}$}{c1,lev1}
      \fmf{fermion,label.side=left,label.dist=.045w,
        label=$q_{i}$}{lev2,lev1}
      \fmf{plain,label.side=right,label.dist=.045w,
        label=$q_{i+1}$}{lev2,cn}
      \fmf{dots}{cn,ifn}
      \fmf{fermion,tension=0,label.side=left,label.dist=.035w,
        label=$k_i$}{lev1,ra1}
      \fmf{fermion,tension=0,label.side=left,label.dist=.035w,
        label=$k_{i+1}$}{ra,lev2}
      \fmfv{decor.size=10,decor.shape=circle,decor.filled=.5}
        {lev1,lev2}
      \fmfv{decor.size=10,decor.shape=circle,decor.filled=0}
        {if1,ifn}
    \end{fmfgraph*}
  \end{center}} 
}


\twocolumn[
\mypreprint{IPPP/07/20\\DCPT/07/40\\LTH 744}
\mytitle{Multijet events in the $\boldmath\rm k_T$-factorisation scheme}
\myauthor{Stefan H{\"o}che$^1$, Frank Krauss$^1$, Thomas Teubner$^2$}
\myinstitute{$^1$ Institute for Particle Physics Phenomenology,
  Durham University, Durham DH1 3LE, UK \\
  $^2$ Department of Mathematical Sciences,
  University of Liverpool, Liverpool L69 3BX, UK}
\myabstract{A Markovian Monte Carlo algorithm for multi-parton production in the
high-energy limit is proposed and the matching with unintegrated
parton densities is discussed.
}]
\mysection{Introduction}
\pagenumbering{arabic}
\setcounter{page}{1}
\label{intro}
Hard scattering at hadron colliders is usually described within the
framework of collinear factorisation
\cite{Collins:2004rq,Collins:1988ig}.  The full scattering
amplitude is factorised into a hard perturbative parton scattering
matrix element and process-independent universal parton distribution
functions (PDFs), which depend on the flavour of the extracted parton, 
its energy or light-cone momentum fraction $x$ w.r.t.\ the initial hadron, 
and on the factorisation scale $\mu_F$.  The choice of this scale is, to
some extent, arbitrary. By the inclusion of higher-order
corrections the dependence of the cross section on this scale is
diminished.  At present, PDFs cannot be obtained from first principles
due to their essentially non-perturbative origin, but they can be
extracted from data, for example through global fits
\cite{Martin:2002aw,Martin:2003sk,Tung:2006tb}.  On the other hand,
the evolution of these collinear PDFs with changing factorisation
scale can be determined perturbatively.  In the collinear
factorisation scheme, all initial state partons are on-shell and have
zero transverse momenta ${\rm k}_\perp=0$.

An alternative approach is the framework of ${\rm k}_\perp$- or
high-energy factorisation.  There, unintegrated PDFs (UPDFs) are convoluted
with off-shell matrix elements.  The PDFs are unintegrated in terms of
the initial partons' ${\rm k}_\perp$.  Initially, ${\rm
k}_\perp$-factorisation has been formulated for heavy quark production
\cite{Catani:1990eg,Collins:1991ty,Levin:1991ya}.  The approach has been 
further investigated in other channels, see for instance
\cite{Hagler:2000eu,Lipatov:2005at,Lipatov:2005tz}.  

The ${\rm k}_\perp$-factorisation has apparent advantages over
conventional collinear factorisation: First, in the high-energy limit,
i.e.\ for $t\ll s$ with $s$ being large, the QCD cross section for jet
production is dominated by gluon exchange diagrams, which diverge in
this limit.  This divergence is alleviated or even removed by realising 
that the $1/t$ divergences in the matrix element can be identified with 
divergences of the form $1/{\rm k}_\perp^2$ and thus using a suitable 
form of unintegrated PDFs, vanishing fast enough for
${\rm k}_\perp\to 0$.  Second, employing UPDFs means including the leading 
logarithmic contribution of higher order corrections to a given process, 
since the effect of additional QCD radiation is encoded in them 
\cite{Ryskin:1999yq,Andersson:2002cf}.
 
Taking the high-energy limit in a given process is equivalent to the
BFKL limit \cite{Kuraev:1977fs,Balitsky:1978ic}, 
which builds on $t$-channel dominance of 
scattering cross sections and the reggeisation of $t$-channel gluons
\cite{Lipatov:1976zz}.  In the past, there have been various
approaches, aiming at a solution of the BFKL dynamics with Monte Carlo
methods and thus producing exclusive final states.  An approximation,
aiming at a correct description of essential features of the BFKL
equation and a correct extrapolation to the DGLAP regime, has been
proposed in the ``Linked Dipole Chain Model''
\cite{Andersson:1995ju,Andersson:1995jt}.  This model has been
implemented in \cite{Kharraziha:1997dn}.  The scope of this approach
is closely related to the CCFM equation
\cite{Ciafaloni:1987ur,Catani:1989yc,Catani:1989sg}. Event
generators based on this evolution equation have been presented in
\cite{Marchesini:1990zy,Jung:2000hk,Jung:2001hx,Golec-Biernat:2007pu}.  
An iterative solution of the pure BFKL equation has been proposed in
\cite{Kwiecinski:1996fm}, iterative Monte Carlo solutions in
\cite{Schmidt:1996fg,Orr:1997im}.  Later on, this prescription has
been extended to next-to-leading logarithmic accuracy 
\cite{Andersen:2003an,Andersen:2003wy,Andersen:2006sp,Andersen:2006kp}.

In this paper, a different implementation of 
${\rm k}_\perp$\--fac\-to\-ri\-sa\-tion for the case of multijet production 
is discussed. Emphasis is put on finding a gauge invariant form of the 
corresponding expressions and on identifying their matching to
unintegrated PDFs derived from conventional collinear
ones\cite{Kimber:1999xc,Kimber:2001sc,Watt:2003mx}.  
It turns out that this in fact
can be achieved by working in the high-energy limit, using as basic
building blocks splitting functions in the limit $z\to 0$,\footnote{In
  addition to the pure gluonic ladders of the high-energy limit, here
  also vertices for quark production are included.}  
in conjunction with a proper reggeisation of all $t$-channel propagators.
Since four-momentum conservation can explicitly be imposed in a Monte
Carlo solution, this approach clearly includes effects beyond the
naive leading order BFKL limit.\footnote{As was discussed for example in 
  \cite{DelDuca:1994ng,Kwiecinski:1996td,Thorne:1999rb,Andersen:2006sp}, 
  the implementation of four-momentum conservation and running $\alpha_s$ 
  effects strongly modifies naive LO BFKL predictions, which were shown 
  to poorly describe data.}  
Furthermore, identifying the probabilistic interpretation of each emission 
in the high-energy limit, the Monte Carlo solution has for the first
time been implemented as a Markovian approach, similar to conventional 
parton shower event generators.  This enables generation of an a priori 
arbitrary number of emissions, which is important at high energies, 
where corrections due to large final state multiplicities are sizable.

The paper is organised as follows: In Sec.~\ref{updfs} the procedure 
of \cite{Kimber:1999xc,Kimber:2001sc,Watt:2003mx,Watt:2003vf} (KMRW) 
to generate doubly unintegrated PDFs (DUPDFs) and the corresponding
angular ordering constraints are reviewed. In Sec.~\ref{dll_proof} 
it is then shown that the leading $\ln(1/x)$ terms are correctly taken
into account.  Section~\ref{mc_procedure} contains the description of
the Markovian MC procedure to generate event topologies with an a
priori undetermined number of final state partons.  In
Sec.~\ref{mc_results} first results are presented and
Sec.~\ref{conclusions} contains our conclusions.


\mysection{Unintegrated parton densities and the KMRW procedure}
\label{updfs}
In this section, the KMRW procedure of constructing unintegrated PDFs
from conventional DGLAP PDFs
\cite{Kimber:1999xc,Kimber:2001sc,Watt:2003mx,Watt:2003vf} is
reviewed.  The discussion and notation closely follows
\cite{Watt:2003mx,Watt:2003vf}.

In collinear factorisation, where the transverse momenta 
${\rm k}_\perp$ of incoming partons are taken to be zero, the parton
densities obey the DGLAP evolution equation
\cite{Altarelli:1977zs,Lipatov:1974qm,Gribov:1972ri,Dokshitzer:1977sg}, 
which determines the $\mu_F$-dependence at fixed light-cone momenta. 
This evolution equation resums leading logarithmic parts of higher 
perturbative orders. In a Monte Carlo formulation, real emission corrections 
can be implemented as a Markov chain of $1\to 2$ parton splittings
\cite{Field:1989uq,Ellis:1991qj,Sjostrand:2003wg}.  However, a study
of QCD beyond double leading logarithmic order reveals that quantum
coherence effects suppress parton emissions in regions of phase space,
where the emission angle of the emitted parton is larger than the
opening angle of the emitting colour dipole 
\cite{Mueller:1981ex,Ellis:1991qj}.  
To exemplify this, consider a parton evolution 
chain in the initial state of a DIS event, as depicted in
Fig.~\ref{fig:angular_ordering_dis}.
\myfigure{t}{
 \angularorderingdis{75}{25}{75}{\normalfeynmf}}{Multiple gluon emission in 
   deep inelastic lepton-nucleon scattering. The hard scattering process is
   characterised by the scale $\mu$.  Usually this scale is also 
   employed as the factorisation scale. \label{fig:angular_ordering_dis}}
If angular ordering is fulfilled, the momenta $k_i$ of the radiated
partons will be distributed such that their angle $\theta_i$ with
respect to the beam direction increases from the incoming proton
towards the hard scattering.  To investigate the implications of this
constraint, it is convenient to start with a Sudakov decomposition of
the momenta \cite{Sudakov:1954sw},
\begin{align}\label{sudakov_decomposition}
  p_i&=x_iP+\tilde\beta_iq'-k_{i\,\perp}\;,&
  k_i&=\alpha_iP+\beta_iq'+k_{i\,\perp}\;,
\end{align}
where $P$ is the proton momentum, $q$ is the photon momentum and
$q'=q+x_B P$, with $x_B$ being the Bj{\o}rken $x$. In the high-energy
limit, the proton mass can be neglected, $m_p^2\ll Q^2=-q^2$. Hence
$q'^2=0$ and in the Breit frame the momenta read
\begin{align*}
  P\,&=\;\frac{1}{2 x_B}\,(Q,{\bf0},Q)\;,\\
  q'\,&=\;\frac{1}{2}(Q,{\bf0},-Q)\quad\text{and}\\
  k_{i\,\perp}\,&=\;(0,{\bf k}_{i\,\perp},0)\;.
\end{align*}
All emitted partons are on-shell, which allows to relate their Sudakov
parameters through
\begin{align*}
  \beta_i\,=\;(\tilde\beta_{i-1}-\tilde\beta_i)\,=\;\frac{z_i}{1-z_i}
  \frac{{\rm k}_{i\,\perp}^2/Q^2}{x_i/x_B}\;,
\end{align*}
where $z_i\,=\,x_i/x_{i-1}$.
Imposing angular ordering for the emissions results in ordering of the
corresponding rapidities $y_i$, since
\begin{equation*}
  y_i=\frac{1}{2}\ln\,\xi_i\,=-\ln\,\tan\frac{\theta_i}{2}\;,
\end{equation*}
where $\xi_i\,=\,k_i^+/k_i^-\,=\,\alpha_i/x_B\,\beta_i$ and $\theta_i$
is the angle of $k_i$ with respect to the beam axis.
According to Eq.~\eqref{sudakov_decomposition}
\begin{equation}\label{emission_rapidity}
  \xi_i=\frac{x_i^2}{x_B^2}
  \left(\frac{1-z_i}{z_i}\frac{Q}{{\rm k}_{i\,\perp}}\right)^2
  =\frac{x_i^2}{x_B^2}\left(\frac{Q}{z_i\,{\rm\bar k}_i}\right)^2\;,
\end{equation}
where the rescaled transverse momentum $\bar{\rm k}_i={\rm
k}_{i\,\perp}/(1-z_i)$ has been introduced.  Hence angular ordering
requirements yield the constraints
\begin{align}\label{rapidity_ordering}
  z_{i}\bar{\rm k}_{i}<\bar{\rm k}_{i+1}\;\;\;\text{and}\;\;z_n\bar{\rm k}_n<\bar{p}\;.
\end{align}
Here $\bar{p}=x_{n+1}Q\sqrt{\Xi}/x_B$ is the maximal rescaled
transverse momentum which is fixed by the hard process through
$\Xi=(1+\tilde\beta_{n+1})/(x_{n+1}/x_B-1)$.  Typically, in an angular
ordered evolution of the parton distributions, $\bar{p}$ plays the
role of the factorisation scale $\mu_F$
\cite{Ciafaloni:1987ur,Catani:1989yc,Catani:1989sg,Marchesini:1994wr}.
The above ordering procedure can be generalised to hadron-hadron
collisions. In this case, both incoming particles have a partonic
substructure.  In general, this leads to two separate factorisation
scales, $\mu_F^{(1)}$ and $\mu_F^{(2)}$, for the two parton densities,
respectively.

In \cite{Kimber:1999xc,Kimber:2001sc,Watt:2003mx,Watt:2003vf} it has
been shown that doubly unintegrated PDFs (DUPDFs) may be inferred from
conventional DGLAP PDFs.  In the following, DUPDFs will be denoted by
$f_a(x,z,{\rm k}_\perp^2,\mu_F^2)$, while their conventional D\-G\-L\-A\-P
counterpart will be denoted by $f_a(x,\mu_F^2)$.  The DU\-PDFs 
must satisfy the normalisation condition
\begin{equation}\label{normalization_dupdf}
  \int_x^1\done z\,\int\frac{\done {\rm k}_\perp^2}{{\rm k}_\perp^2}\,
  f_a(x,z,{\rm k}_\perp^2,\mu_F^2)\,=\;x\,f_a(x,\mu_F^2)\,.
\end{equation}
Employing the Sudakov form factor\footnote{
  The factor of $1/2$ in the sum over the parton species 
  avoids double-counting $s$- and $t$-channel partons.}
\begin{align}\label{sudakov_form_factor}
  &\tilde\Delta_a({\rm k}_\perp^2,\mu_F^2)\\&\quad=
  \;\exp\left\{\,-\int_{{\rm k}_\perp^2}^{\mu_F^2}\frac{\done {\rm k'}_\perp^2}
    {{\rm k'}_\perp^2}\,\frac{\alpha_s({\rm k'}_\perp^2)}{2\pi}\,
    \frac{1}{2}\sum\limits_{b}\,\int_0^1\done\zeta\,
    \tilde{P}_{ab}(\zeta)\,\right\}\;,\nonumber
\end{align}
with $\tilde{P}_{ab}(\zeta)$ denoting regularised DGLAP splitting functions
for the splitting $a\to b$, a singly unintegrated parton distribution 
$\tilde f_a(x,{\rm k}_\perp^2,\mu_F^2)$ is obtained through
\begin{equation}\label{definition_updf}
  \tilde f_a(x,{\rm k}_\perp^2,\mu_F^2)=
    \frac{\partial}{\partial\ln {\rm k}_\perp^2}
    \left[\,x\,f_a(x,{\rm k}_\perp^2)\,
    \tilde\Delta_a({\rm k}_\perp^2,\mu_F^2)\,\right]\;.
\end{equation}
In the region ${\rm k}_\perp^2>\mu_F^2$ this UPDF is set to zero.
This procedure leaves some minimum ${\rm k}_\perp^2$-scale to be
defined, below which DGLAP parton evolution is not valid.  In the
following, this scale will be denoted by $\mu_0^2$.
Relation~\eqref{definition_updf} then holds true only above $\mu_0^2$,
which yields the constraint
\begin{equation*}
  \int_0^{\mu_0^2}\frac{\done {\rm k}_\perp^2}{{\rm k}_\perp^2}\,
  \tilde f_a(x,{\rm k}_\perp^2,\mu_F^2)\,=\;
  x\,f_a(x,\mu_0^2)\,\tilde\Delta_a(\mu_0^2,\mu_F^2)
\end{equation*}
on the singly unintegrated PDF.  Whenever UPDFs,
satisfying this normalisation condition, are applied in 
${\rm k}_\perp$-factorisation, physical observables must be insensitive to
details of the infrared behaviour of $\tilde f_a(x,{\rm k}_\perp^2,\mu_F^2)$, 
i.e.\ below $\mu_0^2$.\footnote{
  It turns out that there is no need for an explicit form of the DUPDFs below 
  $\mu_0^2$, since the $t$-channel parton chains contain a natural cutoff in 
  ${\rm k}_\perp^2$, cf.~\cite{Schmidt:1996fg}, by imposing phase space cuts
  given by physical observables like minijets.}
Therefore, a choice can be made, for example \cite{Watt:2003vf}
\begin{equation*}
  \begin{split}
  &\left.\vphantom{\frac{\partial}{\partial}}
    \tilde f_a(x,z,{\rm k}_\perp^2,\mu_F^2)
    \right|_{\mu_F^2<\mu_0^2}\\
  &\quad=\;
  \frac{{\rm k}_\perp^2}{\mu_0^2}\sbr{A_a\rbr{x,z,\mu_F^2}+
    \frac{{\rm k}_\perp^2}{\mu_0^2}B_a\rbr{x,z,\mu_F^2}}
  \end{split}
\end{equation*}
where
\begin{align*}
A_a\rbr{x,z,\mu_F^2}&=-\tilde{f}_a\rbr{x,z,\mu_0^2,\mu_F^2}\\
  &\quad+\frac{2x}{1-x}\,f_a(x,\mu_0^2)\,\tilde\Delta_a(\mu_0^2,\mu_F^2)\;,\\
B_a\rbr{x,z,\mu_F^2}&=2\tilde{f}_a\rbr{x,z,\mu_0^2,\mu_F^2}\\
  &\quad-\frac{2x}{1-x}\,f_a(x,\mu_0^2)\,\tilde\Delta_a(\mu_0^2,\mu_F^2)\;.
\end{align*}
This choice implies that the UPDF vanishes with ${\rm k}_\perp^2$ for 
${\rm k}_\perp\to 0$, as required by gauge invariance \cite{Gribov:1984tu}.

Instead of the regularised splitting functions $\tilde P_{ab}(z)$,
unregularised splitting functions $P_{ab}(z)$ may safely be used here.
This is because the splitting kernels are implicitly regularised by
imposing the rapidity ordering constraint
Eq.~\eqref{rapidity_ordering}.  Inserting corresponding
$\Theta$-functions in $z$ results in the singly unintegrated quark 
and gluon distributions $f_q(x,{\rm k}_\perp^2,\mu_F^2)$ and 
$f_g(x,{\rm k}_\perp^2,\mu_F^2)$, respectively \cite{Watt:2003mx,Watt:2003vf}.  
The term singly unintegrated indicates that these PDFs depend on
one additional variable w.r.t.\ the collinear ones.  It is
straightforward, however, to introduce an additional $z$-dependence by
simply dropping the $z$-integration in
Eq.~\eqref{normalization_dupdf}.  Such defining the DUPDF
\begin{equation}\label{definition_dupdf}
  \begin{split}
    &f_a(x,z,{\rm k}_\perp^2,\mu_F^2)\\
    &\quad=\;\Delta_a({\rm k}_\perp^2,\mu_F^2)\,
      \frac{\alpha_s({\rm k}_\perp^2)}{2\pi}\,\sum\limits_{b}\,P_{ba}(z)\,
      \frac{x}{z}\,f_b\left(\frac{x}{z},{\rm k}_\perp^2\right)\,\\
      &\quad\quad\quad\times\sbr{\,(1-\delta_{ab})+\delta_{ab}\,\Theta
         \left(\frac{\textstyle\mu_F}{\textstyle\mu_F+{\rm k}_\perp}-z\right)\,}
  \end{split}
\end{equation} 
the desired relation, Eq.~\eqref{normalization_dupdf}, is
satisfied for both parton species.  To guarantee the consistency of
the approach, the conventional DGLAP PDF employed to obtain the DUPDFs
should be determined using the leading order unregularised
splitting kernels employed in Eq.~\eqref{definition_dupdf}.
Furthermore, a consistent treatment of the running coupling $\alpha_s$
should be imposed.


\mysection{DUPDFs as impact factors for LL BFKL evolution}
\label{dll_proof}
In this section we argue that the DUPDFs defined above may be employed
as impact factors in the calculation of multi-gluon cross sections in
the high-energy limit.  The argument works at leading logarithmic (LL)
accuracy.  The starting point is the integrated LL gluon branching 
probability $\Gamma^{(LL)}_g=-\log\Delta^{(LL)}_g$, which determines 
the behaviour of the DGLAP evolution of the gluon density.\footnote{
  The factor $1/2$ contained in Eq.~\eqref{sudakov_form_factor} 
  must be cancelled here in order to restore the $t$/$u$-symmetry 
  of the splitting process.}
\begin{equation*}
  \Gamma^{(LL)}_g(\mu^2,\tilde\mu^2)\,=\,
    \Gamma^{(LL)}_{gg}(\mu^2,\tilde\mu^2)+\sum\limits_{q}
    \Gamma^{(LL)}_{gq}(\mu^2,\tilde\mu^2)\;,\\
\end{equation*}
where
\begin{equation*}
  \Gamma^{(LL)}_{ab}(\mu^2,\tilde\mu^2)\,=\,
  \int_{\,\ln\mu^2}^{\,\ln\tilde\mu^2}\done\ln {\rm k}_\perp^2
  \int^{\frac{\tilde\mu}{\tilde\mu+{\rm k}_\perp}}_
  {\frac{{\rm k}_\perp}{\tilde\mu+{\rm k}_\perp}}\done z\,
  \frac{\alpha_s}{2\pi}\,P_{ab}(z)\;,
\end{equation*}
with $P_{ab}(z)$ again denoting the unregularised 
DGLAP splitting kernels and the integration boundaries 
determined by angular ordering, cf.\ the $\Theta$-function 
in Eq.~\eqref{definition_dupdf}. To simplify the discussion we firstly
focus on $\Gamma^{(LL)}_{gg}$ only.
The corresponding part of the Sudakov form factor reads
\begin{equation*}
  \Delta^{(LL)}_{gg}(\mu^2,\tilde\mu^2)=\exp\left\{
  -\Gamma^{(LL)}_{gg}(\mu^2,\tilde\mu^2)\right\}\;.
\end{equation*}
Replacing the splitting variable $z$ of the emitter parton by the
rapidity $y$ of the emission, which, according to
Eq.~\eqref{emission_rapidity} is given by
\begin{equation*}
  y\,=\;\frac{1}{2}\ln\xi\,=\;\ln\left(\frac{x}{x_B}
  \frac{Q}{{\rm k}_\perp}\right)-\ln\frac{z}{1-z}\;
\end{equation*} 
results in
\begin{equation}\label{incomplete_dglap_sudakov_pre}
  \begin{split}
  &\Gamma^{(LL)}_{gg}(\mu^2,\tilde\mu^2)\,\\&\quad=\,-
  \int_{\,\ln\mu^2}^{\,\ln\tilde\mu^2}\done\ln {\rm k}_\perp^2
  \int_{y\left(z_{\rm min}\right)}^{y\left(z_{\rm max}\right)}
  \done y\\&\quad\quad\quad\times\frac{2C_A\,(1-z(1-z))^2}{P_{gg}(z)}\,
  \frac{\alpha_s}{2\pi}\,P_{gg}(z)\\&\quad=\;
  \int_{\,\ln\mu^2}^{\,\ln\tilde\mu^2}\done\ln {\rm k}_\perp^2
  \int^{y\left(z_{\rm min}\right)}_{y\left(z_{\rm max}\right)}\done y\;
  \tilde{\alpha}_s\;(1-z(1-z))^2\;,
  \end{split}
\end{equation}
where $\tilde{\alpha}_s=\alpha_s C_A/\pi$.  The term $z(1-z)$ in the
numerator corresponds to helicity non-conserving configurations in the
$1\to 2$ parton splittings and thus in the impact
factor~\cite{DelDuca:1999ha}. These configurations are absent
in the high-energy limit, which simplifies the integrand of 
Eq.~\eqref{incomplete_dglap_sudakov_pre}, such that the part of the 
integrated LL gluon branching probability induced by $g\to gg$ 
splittings reads
\begin{equation}\label{incomplete_dglap_sudakov}
  \begin{split}
  \Gamma^{(LL)}_{gg}(\mu^2,\tilde\mu^2)\,=\;&
  \int_{\,\ln\mu^2}^{\,\ln\tilde\mu^2}\done\ln {\rm k}_\perp^2
  \int^{y\left(z_{\rm min}\right)}_{y\left(z_{\rm max}\right)}\done y\;
  \tilde{\alpha}_s\;.
  \end{split}
\end{equation}
Keeping in mind that $\tilde{\alpha}_s$ depends on transverse degrees 
of freedom only, performing the $y$-integration results in
\begin{equation*}
  \begin{split}
  &\Gamma^{(LL)}_{gg}(\mu^2,\tilde\mu^2)\,\\&\quad=\,
  \int_{\,\ln\mu^2}^{\,\ln\tilde\mu^2}\done\ln {\rm k}_\perp^2\,
  \tilde{\alpha}_s\;\\&\quad\quad\quad\times
  \cbr{\ln\rbr{\frac{\tilde\mu}{{\rm k}_\perp}\frac{x Q}{x_B {\rm k}_\perp}}-
    \ln\rbr{\frac{{\rm k}_\perp}{\tilde\mu}\frac{x Q}{x_B {\rm k}_\perp}}}\\
  &\quad=\frac{1}{2}\int_{0}^{\,\ln^2\tilde\mu^2/\mu^2}\done\ln^2
  \frac{\tilde\mu^2}{{\rm k}_\perp^2}\;\tilde{\alpha}_s\;.
  \end{split}
\end{equation*}
The order of integration in Eq.~\eqref{incomplete_dglap_sudakov} 
may be changed,
\begin{equation}\label{modified_dglap_sudakov}
  \begin{split}
    \Gamma^{(LL)}_{gg}(\mu^2,\tilde\mu^2)\,=&\;
    \int_{y}^{\tilde y}\done y'\,\int_{0}^{\,\ln\tilde\mu^2/\mu^2}
    \done\ln\frac{\tilde\mu^2}{{\rm k}_\perp^2}\;\tilde{\alpha}_s\;\\
    &\quad\times\Theta\rbr{\,\ln\tilde\mu^2/{\rm k}_\perp^2+y-y'\,}\;,
  \end{split}
\end{equation}
where $\tilde y=\ln x/x_B+\ln Q/\tilde\mu$ and
$\tilde{y}-y=\ln\tilde\mu^2/\mu^2$.

If the running coupling is treated identically, this result agrees
with the reggeisation factor of the $t$-channel gluon propagator found
by rewriting Eq.~(7) of~\cite{Schmidt:1996fg}. Up to a minor 
transformation, this equation reads\footnote{Note that
  the particle indices $a$ and $b$ are interchanged with respect 
  to Schmidt's original formulation.}
\begin{align}\label{n_gluon_exchange_me}
  &f^n\rbr{y_{ab},\,{\rm p}_{a\perp},\,{\rm p}_{b\perp}}\\
  &\nonumber\quad=\;\int\prod_{i=1}^n\sbr{\bar\alpha_s
    \done y_i\frac{\done{\rm k}_{i\perp}^2}{{\rm k}_{i\perp}^2}
    \frac{\done\phi_i}{2\pi}\exp\cbr{\;
      -\bar\alpha_s\ln\frac{{\rm q}_{i\perp}^2}{\mu_0^2}\Delta y_i}}\\
  &\nonumber\quad\quad\quad\times\exp\cbr{-\bar\alpha_s\ln
    \frac{{\rm q}_{0\perp}^2}{\mu_0^2}\Delta y_0}
    \frac{1}{2}\delta\rbr{{\rm p}_{b\perp}+{\rm q}_{n\perp}}\,,
\end{align}
where $\bar{\alpha}_s = \alpha_sC_A/\pi$ and $q_i=p_a-\sum_{j=1}^i k_j$.
The exponential term in the square brackets is readily identified as
\begin{equation}\label{tchannel_reggeisation}
  \bar\Delta(y,\tilde y)=\exp\left\{
  -\bar\Gamma^{(LL)}_g(y,\tilde y)\right\}\;,
\end{equation}
where
\begin{equation*}
  \bar\Gamma^{(LL)}_g(y,\tilde y)=\int_y^{\tilde y}
  \done y'\int^{\,\ln{\rm q}_\perp^2/\mu_0^2}_{0}
  \done\ln \frac{{\rm q}_\perp^2}{{\rm k}_\perp^2}\;\bar{\alpha}_s\;,
\end{equation*}
which is the desired result. It has been pointed out 
e.g.\ in~\cite{Andersson:2002cf} that the comparison with NLO BFKL 
calculations suggests the choice $\alpha_s = \alpha_s({\rm k}_{\perp}^2)$, 
similar to the DGLAP case. Employing
\begin{equation*}
  \alpha_s({\rm k}_{\perp}^2)=\frac{1}{\beta_0
    \log {\rm k}_{\perp}^2/\Lambda^2}\;,\quad{\rm where}\quad
  \beta_0=\frac{11-2/3N_f}{4\pi}\;,
\end{equation*}
we then end up with the result presented in~\cite{Orr:1997im},
\begin{equation*}
  \bar\Gamma^{(LL)}_g(y,\tilde y)=\rbr{\tilde y-y}\,
    \frac{C_A}{\pi\beta_0}\,
    \log\rbr{\frac{\alpha_s(\mu_0^2)}{\alpha_s(q_\perp^2)}}\;.
\end{equation*}
In our numerical analyses, $\Lambda$ is chosen consistent 
with the input PDF. Equation~\eqref{n_gluon_exchange_me}
can be used to construct the full LL BFKL kernel $f$ through
\begin{equation}\label{bfkl_kernel_mc}
  f\rbr{y_{ab},\,{\rm p}_{a\perp},\,{\rm p}_{b\perp}}\,=\;
  \sum\limits_{n=0}^\infty\,
    f^n\rbr{\,y_{ab},\,{\rm p}_{a\perp},\,{\rm p}_{b\perp}}\;.
\end{equation}
Since rapidity ordering is trivially satisfied in the BFKL evolution,
the explicit ordering requirement incorporated in the $\Theta$-function
of Eq.~\eqref{modified_dglap_sudakov} may be dropped whenever
$\bar\Delta(y,\tilde y)$ is employed.

Following the same reasoning, $\Gamma^{(LL)}_{gq}$ is given by
\begin{equation*}
  \begin{split}
    &\Gamma^{(LL)}_{gq}(\mu^2,\tilde\mu^2)\,\\&\quad=\;
    \int_{\,\ln\mu^2}^{\,\ln\tilde\mu^2}\done\ln {\rm k}_\perp^2
    \int^{y\left(z_{\rm min}\right)}_{y\left(z_{\rm max}\right)}\done y\;
    \tilde{\alpha}_s\,\\&\quad\quad\quad\times
    \frac{T_R}{C_A}\;\frac{1}{2}z(1-z)\rbr{\,z^2+(1-z)^2}\;.
  \end{split}
\end{equation*}
In principle, this term vanishes in the high-energy limit
due to the prefactor $z(1-z)$, thus allowing to identify 
$\bar\Delta(y,\tilde y)$ with $\Delta^{(LL)}_g(\mu^2,\tilde\mu^2)$.  
However, it may be used to model quark production along the BFKL ladder, 
as will be discussed in Sec.~\ref{mc_quarks}.

Similar considerations may be applied to the integrated quark branching
probability.  Starting from the expression
\begin{equation*}
  \Gamma^{(LL)}_q(\mu^2,\tilde\mu^2)\,=\,
    \Gamma^{(LL)}_{qg}(\mu^2,\tilde\mu^2)+
    \Gamma^{(LL)}_{qq}(\mu^2,\tilde\mu^2)
\end{equation*}
and again replacing the splitting variable $z$ by the rapidity $y$ 
results in
\begin{equation*}
  \begin{split}
  &\Gamma^{(LL)}_{qg}(\mu^2,\tilde\mu^2)\,\\&\quad=\;
    \int_{\,\ln\mu^2}^{\,\ln\tilde\mu^2}\done\ln {\rm k}_\perp^2
    \int^{y\left(z_{\rm min}\right)}_{y\left(z_{\rm max}\right)}\done y\;\\
  &\quad\quad\quad\times
    \tilde{\alpha}_s\,\frac{C_F}{2C_A}\;\rbr{1-z}\rbr{\,1+(1-z)^2}\;.
  \end{split}
\end{equation*}
By identifying $z=-t/s$, all factors $1-z$ become unity in
the high-energy limit.  Thus,
\begin{align*}
  \Gamma^{(LL)}_{qg}(\mu^2,\tilde\mu^2)\,=\;&
  \frac{C_F}{C_A}\,\Gamma^{(LL)}_{gg}(\mu^2,\tilde\mu^2)\;.
\end{align*}
Simultaneously, due to the denominator part $(1-z)$ in $P_{qq}(z)$
quark production in the $t$-channel is suppressed,
hence allowing to identify
\begin{align*}
  \Gamma^{(LL)}_{q}(\mu^2,\tilde\mu^2)\,=\;&
  \frac{C_F}{C_A}\,\Gamma^{(LL)}_{g}(\mu^2,\tilde\mu^2)\;.
\end{align*}
However, $\Gamma^{(LL)}_{qq}(\mu^2,\tilde\mu^2)$ may be employed
to model gluon emission from $t$-channel quark lines, as will be
described in Sec.~\ref{mc_quarks}.

The above considerations show that to leading logarithmic accuracy 
the DUPDFs, Eq.~\eqref{definition_dupdf}, resemble all features of the BFKL
evolution.  Therefore, they can safely be employed as impact factors
for the calculation of cross sections in the high-energy limit.

\mysection{Markovian Monte Carlo solution
  to the $\bf ln(1/x)$-evolution}
\label{mc_procedure}
The Markovian approach to the calculation of cross sections and
differential distributions in the high-energy limit will be presented
in this section.  The advantage of the algorithm is that the number of
emissions stays a priori undetermined, similar to the case of
conventional parton showers employed to solve $\log(Q^2/\mu^2)$-evolution 
\cite{Field:1989uq,Ellis:1991qj,Sjostrand:2003wg}.  
The factorisation of the radiation pattern into individual emissions, 
which depend on each other merely through the correct ordering, 
allows to model further physics effects involving the produced
outgoing partons, like for instance adding final state radiation.

The basis of the formalism is encoded in Eq.~(7) in
\cite{Schmidt:1996fg} and Eq.~\eqref{tchannel_reggeisation}.  These
equations translate into the probability for having an additional
emission from the BFKL kernel being approximately distributed
according to the function
\begin{equation}\label{y_selection_rule}
  \begin{split}
  &\gamma\rbr{1,\Gamma_g^{(LL)}(y_i,y_n)}\,\\&\quad=\;
  \Gamma_g^{(LL)}(y_i,y_n)\,\exp\left\{-\Gamma_g^{(LL)}(y_i,y_n)\right\}\;.
  \end{split}
\end{equation}
Here, $y_i$ is the rapidity of the previous and $y_n$ is the rapidity
of the final emission.  Such distributions may be generated employing
the veto algorithm, described for example in \cite{Sjostrand:2003wg}.  
It allows to simultaneously select the rapidity and transverse momentum
of the new emission.\footnote{ In fact applying a veto is not
necessary here, as long as quark production is neglected in the
approach.}  In the following, the superscripts $^{(LL)}$ will be
dropped.

To determine the corresponding $z$-\-${\rm k}_\perp$-\-factorisation
for\-mu\-la, the simplest case, a gluon ladder with no emission, is
investigated.  This corresponds to a ``$2\to 0$ process'' in the
$z$-\-${\rm k}_\perp$-\-factorisation approach.  When working in
collinear factorisation rather than with the DUPDF prescription of
\cite{Watt:2003vf}, it is a $2\to 2$ process.  
The corresponding phase space element can thus be determined by
factorising the collinear matrix element and its phase space
integral. The starting point is
\begin{equation}\label{sigma_dglap_twojet}
  \begin{split}
    \sigma\,=&\;\sum\limits_{a^{(1)},a^{(2)}}
      \int\done \xi^{(1)}\int\done \xi^{(2)}
      \int\frac{\done^4 k_1}{(2\pi)^3}\int\frac{\done^4 k_2}{(2\pi)^3}\;\\
    &\times\delta\rbr{k_1^2}\,\delta\rbr{k_2^2}
      (2\pi)^4\delta^{(4)}(P-k_1-k_2)\,\\ 
    &\times f_{a^{(1)}}(x^{(1)},Q^2)\,f_{a^{(2)}}(x^{(2)},Q^2)\,
    \frac{\left|M_{a^{(1)}\,a^{(2)}}\right|^2}
      {2\,\xi^{(1)}\xi^{(2)}S}\frac{1}{2}\;,
  \end{split}
\end{equation}
where the factor $1/2$ is due to the identity of the final state
particles, $Q^2$ denotes the factorisation scale, $P^2=s$, 
$s=\xi^{(1)}\xi^{(2)}S$, $\xi=x/z$, and the
superscripts $^{(1)}$ and $^{(2)}$ refer to the left and right beam,
respectively.  The matrix element reads
\begin{equation}\label{four_gluon_me}
  \left|M_{gg}\right|^2\,=\;(4\pi\alpha_s)^2\frac{C_A^2}{2}
  \left(3-\frac{tu}{s^2}-\frac{us}{t^2}-\frac{st}{u^2}\right)\;.
\end{equation}
Employing $z_1=z_2=z$, $t=-zs$ and $u=-(1-z)s$ transforms this into
\begin{equation*}
  \left|M_{gg}\right|^2\,=\;(4\pi\alpha_s)^2
  \frac{1}{8}\left[\,P_{gg}(z)\,\right]^2
  \left\{\,1+\mathcal{O}\left(z(1-z)\right)\,\right\}
\end{equation*}
where terms proportional to $z(1-z)$ in the numerator vanish in the
high-energy limit and are not explicitly displayed.

The phase space element of the general case of a gluon ladder with an
arbitrary number of gluons emitted between the two outermost jets can
be derived by combining their momenta into one final state momentum
$K$.  Ignoring the substructure of $K$, the differential two-particle
initial and final state phase space element for the remaining degrees
of freedom reads
\begin{equation*}
  \begin{split}
  \done\Phi_2=\;&\done\xi^{(1)}\done\xi^{(2)}\,
    \frac{\dfour k_1}{(2\pi)^3}\,\frac{\dfour k_2}{(2\pi)^3}\,
    \delta\rbr{k_1^2}\,\delta\rbr{k_2^2}\,\\
    &\times(2\pi)^4\,\delta^{(4)}\rbr{P-K-k_1-k_2}\,,
  \end{split}
\end{equation*}
with $P$ again the total four momentum of the process.  Employing the
four-dimensional $\delta$-function and the relations
$\done\xi^{(1)}\done\xi^{(2)}=\done y\done s/S$ and $\done
p_z=\,\done\rbr{\sqrt{s_\perp}\sinh y} =\,\sqrt{s_\perp}\cosh y\,\done
y=\,E\,\done y$ results in
\begin{equation*}
  \done\Phi_2=\frac{2\pi}{S}\,\done s\,\done y\,
  \frac{\done y_1\,\done {\rm k}_{1\perp}^2\done\phi_1}{4(2\pi)^3}\,
  \delta\rbr{(P-K-k_1)^2}\;.
\end{equation*}
Furthermore, the definition $\bar{P}=P-k_2$ allows to rewrite
\begin{equation*}
  \begin{split}
    \tder{y}{y_2}&=\,\tder{}{y_2}\,\frac{1}{2}\ln
    \frac{\bar{P}^++m_{2\perp}e^{+y_2}}
      {\bar{P}^-+m_{2\perp}e^{-y_2}}\\
    &=\,\frac{1}{2}\rbr{\frac{m_{2\perp}e^{+y_2}}{P^+}+
      \frac{m_{2\perp}e^{-y_2}}{P^-}}
    =\,\frac{Pk_2}{s}\;.
  \end{split}
\end{equation*}
Using $P=\sqrt{s}\,(\cosh{y},\vec{0},\sinh{y})$ gives
\begin{equation*}
  \begin{split}
  &\done s\,\delta\rbr{s+K^2-2P(K+k_1)+2Kk_1}\\&\quad=
  \frac{s}{s-P(K+k_1)}=\frac{s}{Pk_2}\;,
  \end{split}
\end{equation*}
such that
\begin{equation*}
  \begin{split}
    \done\Phi_2
    =&\frac{1}{4S\,(2\pi)^2}\;\done y_2\,
    \done y_1\,\done {\rm k}_{1\perp}^2\done\phi_1\,.
  \end{split}
\end{equation*}

Finally, when fixing the factorisation scale in 
Eq.~\eqref{sigma_dglap_twojet} and the renormalisation scale 
in Eq.~\eqref{four_gluon_me} to be the transverse momentum 
in the process and adding a Regge suppression factor for
the $t$-channel gluon, the $z$-\-${\rm k}_\perp$-\-factorisation 
formula reads
\begin{equation}\label{z_kperp_factorisation_lowxgg}
  \begin{split}
    \sigma\,=&\;\frac{\pi^2}{2S}
    \int\done y_1\,
    \int\done {\rm k}_{1\perp}^2\,\int\done\phi_1\,
    \int\done y_2\,\\
    &\times\,
    \bar f_g(x^{(1)},z,{\rm k}_{\perp}^2,\bar{\rm k}_{\perp}^2)\,
    \bar f_g(x^{(2)},z,{\rm k}_{\perp}^2,\bar{\rm k}_{\perp}^2)\,\\
    &\times\,
    \frac{1}{2\,\xi^{(1)\,2}\xi^{(2)\,2}S}\,
    \frac{1}{\bar\Delta_g(y_1,y_2)}\;.
  \end{split}
\end{equation}
Here, $\bar f_g$ is defined such that only gluon splittings
are contained in the sum over parton species of
Eq.~\eqref{definition_dupdf} and angular ordering is implemented by
the DUPDFs, while $\bar\Delta_g(y_1,y_2)$ is given by 
Eq.~\eqref{tchannel_reggeisation}. The superscripts $^{(1)}$ and $^{(2)}$ 
refer to the left and right beam, respectively. Since the emitted
gluons are distinguishable due to rapidity ordering, the symmetry
factor $1/2$ appearing in Eq.~\eqref{four_gluon_me} must be dropped.
The factorisation scale $\mu_F$ of each DUPDF introduced in 
Eq.~\eqref{definition_dupdf} is unambiguously determined by the 
rescaled transverse momentum $\bar{\rm k}_{\perp}$ of the emissions.

Equation~\eqref{z_kperp_factorisation_lowxgg} describes a gluon ladder
with no rung, but it can be easily extended to final states with an 
arbitrary number of gluons. In contrast to the previous case, the 
momentum fractions $z^{(1)}$ and $z^{(2)}$ are then generally different
from each other. Hence we define the rescaled transverse momenta
$\bar{\rm k}_{2\perp}^{(1)}={\rm k}_{2\perp}/(1-z^{(1)})$ and 
$\bar{\rm k}_{n-1\perp}^{(2)}={\rm k}_{n-1\perp}/(1-z^{(2)})$. 
Employing Eq.~(7) of~\cite{Schmidt:1996fg}, the cross section for 
the $2\to n$ gluon scattering reads
\begin{align}\label{z_kperp_factorisation_lowxng}
  &\sigma\,=\;\frac{\pi^2}{2S}
  \int\done y_1\,\int\done {\rm k}_{1\perp}^2\int
  \done\phi_1\,\int\done y_n\,\\&\nonumber\times\,
  \bar f_g(x^{(1)},z^{(1)},{\rm k}_{1\perp}^{2},
    \bar{\rm k}_{2\perp}^{(1)2})\,
  \bar f_g(x^{(2)},z^{(2)},{\rm k}_{n\perp}^{2},
    \bar{\rm k}_{n-1\perp}^{(2)2})\,\\&
  \nonumber\times\frac{1}{2\,\xi^{(1)\,2}\xi^{(2)\,2}S}\,
  \frac{1}{\bar\Delta_g(y_1,y_2)}\,\left[\;
    \prod\limits_{i=2}^{n-1}
    \int\frac{\done\phi_i}{2\pi}\,\right.\\
  &\nonumber\quad\times\,\left.\int\done y_i
  \int\frac{\done{\rm k}_{i\perp}^2}{{\rm k}_{i\perp}^2}
  \frac{\alpha_s({\rm k}_{i\perp}^2)}{\pi}\,C_{gg}\,
  \bar\Delta_g(y_i,y_{i-1})\;\right]\;,
\end{align}
where
\begin{equation*}
  C_{gg}=C_A\;.
\end{equation*}

The corresponding Monte Carlo event generation algorithm 
can be described as follows:
\begin{enumerate}
\item Determine the kinematics of the first emission and the
  rapidity of the last emission according to the modified
  $z$-$k_\perp$-factorisation formula, 
  Eq.~\eqref{z_kperp_factorisation_lowxng}.
\item As long as phase space allows, choose a new rapidity $y_i$ 
  according to Eq.~\eqref{y_selection_rule} and a new
  transverse momentum ${\rm k}_{i\perp}$. The corresponding cuts on
  the individual emissions have already been discussed in
  \cite{Schmidt:1996fg}.  In the notation employed ibidem, 
  they are given by
  \begin{equation*}
    {\rm k}_{i\perp}^2>\mu_0^2\quad\text{and}\quad
    {\rm q}_{i\perp}^2>\mu_0^2\;.
  \end{equation*}
\item Fix the transverse momentum of the last emission\\ 
  through overall momentum conservation.
\end{enumerate}

\mysection{Model for quark production}
\label{mc_quarks}
So far, it has been shown that
Eq.~\eqref{z_kperp_factorisation_lowxng} yields the correct LL gluon
evolution in the high-energy limit.  In this limit quark production is
strongly suppressed due to the spin structure entering the corresponding
vertices.  However, energies and rapidity intervals at real
colliders are finite and quarks do appear as final state
partons.  Since, for instance, heavy quark production is of large
phenomenological interest, it needs to be described.  In our approach
we aim at not spoiling the high-energy gluon evolution.  Therefore we
choose to model quark production within the BFKL ladder structure by
simply adding a $g^*q^*\to q$ effective vertex, which vanishes in the
high-energy limit, but keeping the finite, non-leading terms.
Additionally, quarks can be produced by employing $qg^*\to q$ and
$qq^* \to g$ impact factors contained within the DU\-PDFs. These
quarks may further radiate gluons, which is modelled by a $q^*q^*\to g$ 
vertex. Figure~\ref{fig:lo_levs_quarks} shows a possible configuration
of quark production.
\myfigure{t}{
  \levs{75}{20}{75}{\normalfeynmf}}{Multi-Regge amplitude
  including the emission of a quark pair with the particle indices 
  $i$ and $i+1$. The shaded blobs represent the vertices proposed
  in Eq.~\eqref{qg_lev}.\label{fig:lo_levs_quarks}}

Following Sec.~\ref{dll_proof}, the remaining vertices are then 
readily determined. At leading logarithmic accuracy they are given by
the corresponding DGLAP splitting functions in the high-energy limit,
\begin{equation}\label{qg_lev}
  \begin{split}
    C_{qg}&=C_F\;,\\
    C_{qq}(z_i)&=\frac{1}{2}C_F\,z_i\;,\\
    C_{gq}(z_i)&=\frac{1}{2}T_R\,z_i\;.
  \end{split}
\end{equation}
Then, the general case of a parton cascade in the high-energy limit reads
\begin{align}\label{z_kperp_factorisation_lowx}
  &\sigma\,=\;\frac{\pi^2}{2S}\sum\limits_{a^{(1)}}
  \int\done y_1\int\done {\rm k}_{1\perp}^2
  \int\done\phi_1\int\done y_n\,\\
  \nonumber\times&\,
  f^{(1)}(x^{(1)},z^{(1)},{\rm k}_{1\perp}^{2},
    \bar{\rm k}_{2\perp}^{(1)2})\,
  f^{(2)}(x^{(2)},z^{(2)},{\rm k}_{n\perp}^{2},
    \bar{\rm k}_{n-1\perp}^{(2)2})\,\\
  \nonumber\times&\frac{1}{2\xi^{(1)\,2}\xi^{(2)\,2}S}
    \frac{1}{\Delta_{a_1}(y_1,y_2)}\;
  \,\left[\;\prod\limits_{i=2}^n
  \int\frac{\done\phi_i}{2\pi}\,\int\done y_i
  \int\frac{\done{\rm k}_{i\perp}^2}{{\rm k}_{i\perp}^2}\right.\\
  \nonumber&\quad\times\left.
  \frac{\alpha_s({\rm k}_{i\perp}^2)}{\pi}\sum\limits_{a_i}
  C_{a_{i-1} a_i}({\rm q}_{i-1},{\rm k}_i)\,
  \Delta_{a_i}(y_i,y_{i-1})\;\right]\;,
\end{align}
where now both quarks and gluons are contained in the sums over parton
species.

If heavy quarks are included in the simulation, their masses are taken
care of in the Reggeisation factor and the phase space
integration. Following the discussion in \cite{Rodrigo:2003ws}, the
branching probability $\Gamma^{(LL)}_Q(y,\tilde{y})$ for
heavy quarks of mass $m$ is modified by
\begin{equation}\label{massive_quark_splitting}
  C_{qg} \quad\longrightarrow\quad 
  \frac{{\rm k}_\perp^2}{{\rm k}_\perp^2+m^2}\,C_{qg}\;.
\end{equation}
Accordingly all external momenta are constructed employing the correct
on-shell masses of the corresponding particles.

\mysection{Results}
\label{mc_results}
In this section, results obtained with the Monte Carlo algorithm 
described above will be presented. All of them have been obtained 
with an implementation into the MC event generator 
Sherpa~\cite{Gleisberg:2003xi}.\footnote{This code is available 
  from the authors upon request.} To eliminate possible dependencies on the
phase space integration, we have cross-checked our calculations with a
different integration method.  This method uses an iterative approach
to generate event topologies for a fixed number of final state particles, 
as explained in the appendix. We found no deviations from our results
generated in the Markovian approach.
\myfigure{t!}{
  \includegraphics[width=\textwidth/2-5mm]{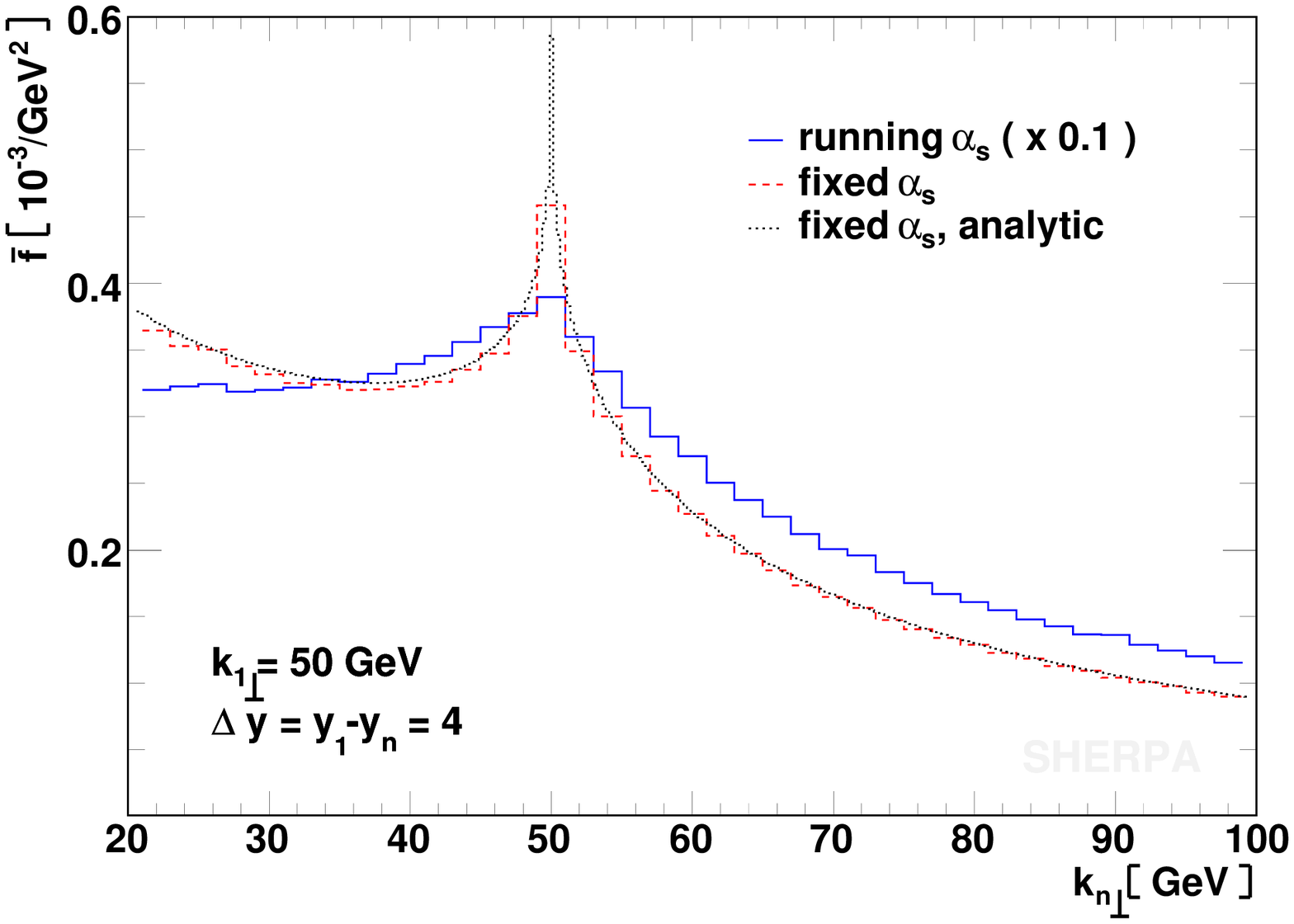}
  }{Transverse momentum spectra $\bar{f}\rbr{{\rm k}_{n\perp}}$ for fixed and
    running coupling solution of Eq.~\eqref{z_kperp_factorisation_lowxng} 
    at fixed ${\rm k}_{1\perp}$ and $\Delta y$. Note that the result
    for running coupling has been rescaled by a factor of 0.1.
    \label{bfkl_pt_b}} 

Firstly, we focus on purely gluonic processes, reflecting the
behaviour of the LO BFKL equation. This essentially translates into
invoking Eq.~\eqref{z_kperp_factorisation_lowxng} for event generation. 
In Fig.~\ref{bfkl_pt_b} the azimuthally averaged ${\rm k}_{n\perp}$ 
spectrum $\bar{f}\rbr{{\rm k}_{n\perp}}=\abr{f\rbr{{\rm k}_{n\perp}}}_\phi$ 
is shown, where we have fixed ${\rm k}_{1\perp}$ = 50~GeV and $\Delta y$ = 4, 
and where the DUPDFs have been set to 1.
Therefore, this plot investigates the behaviour of the BFKL kernel,
Eq.~\eqref{bfkl_kernel_mc}, only. As collider setup, the LHC
with a c.m.\ energy of 14~TeV has been chosen. In the fixed coupling 
solution $\alpha_s$ has been evaluated at scale ${\rm k}_{1\perp}^2$.  
The figure shows the effect of going from a fixed coupling and 
unconstrained kinematics to a running coupling with kinematical
constraints, which considerably widens the distribution. Also, since
$\alpha_s$ is typically evaluated at smaller scales, $\bar{f}$ is
significantly enhanced.  The large influence of kinematical
constraints and running coupling on the BFKL dynamics has already been
noted, e.g.\ in \cite{Andersen:2006sp,Thorne:1999rb}.
\myfigure{t}{
  \includegraphics[width=\textwidth/2-5mm]{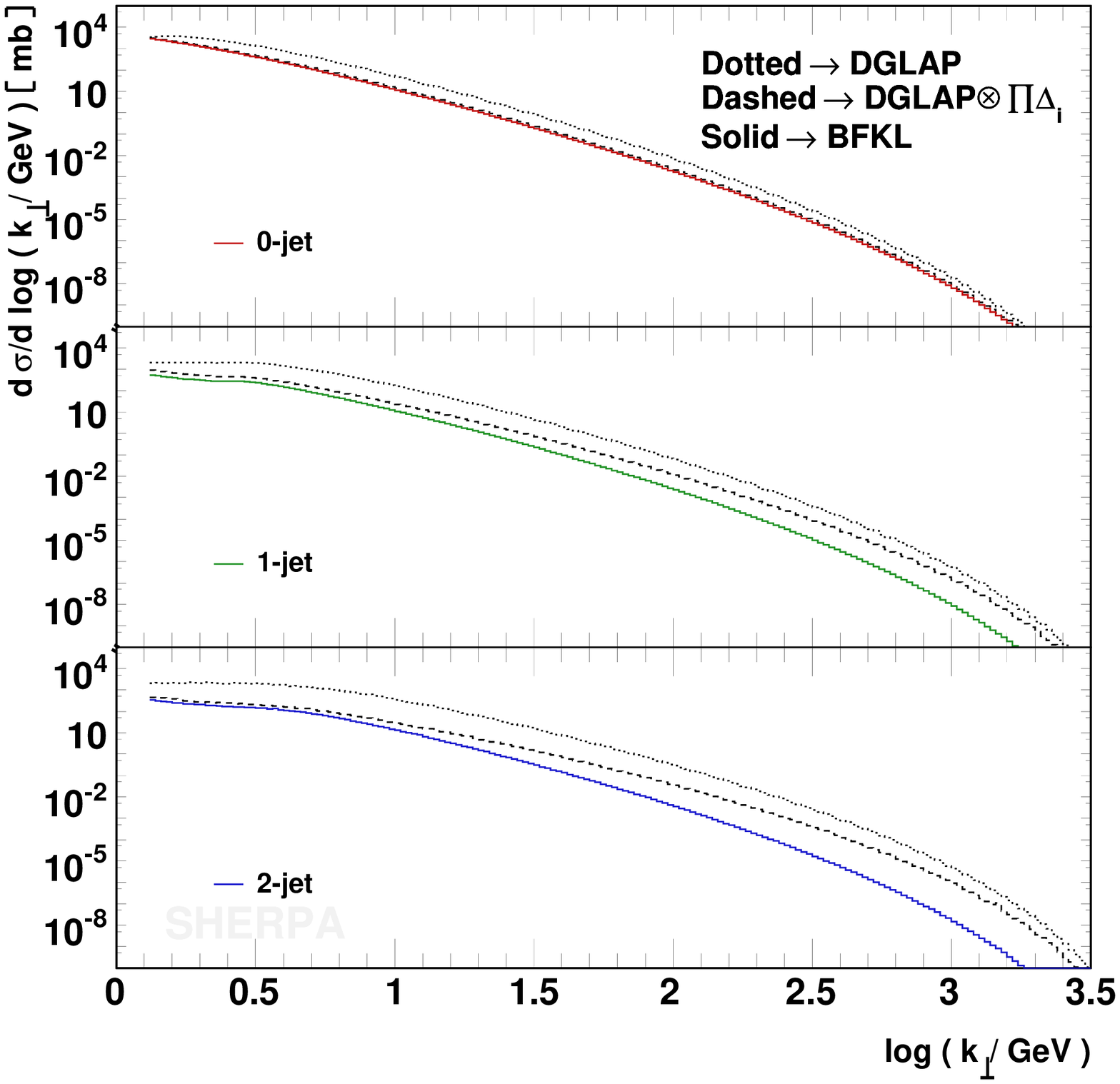}
  }{Comparison of $\log\rbr{{\rm k}_\perp}$-distributions between BFKL 
    and reweighted DGLAP matrix elements.\label{dglap_bfkl_ikt}}
\myfigure{b!}{
  \includegraphics[width=\textwidth/2-5mm]{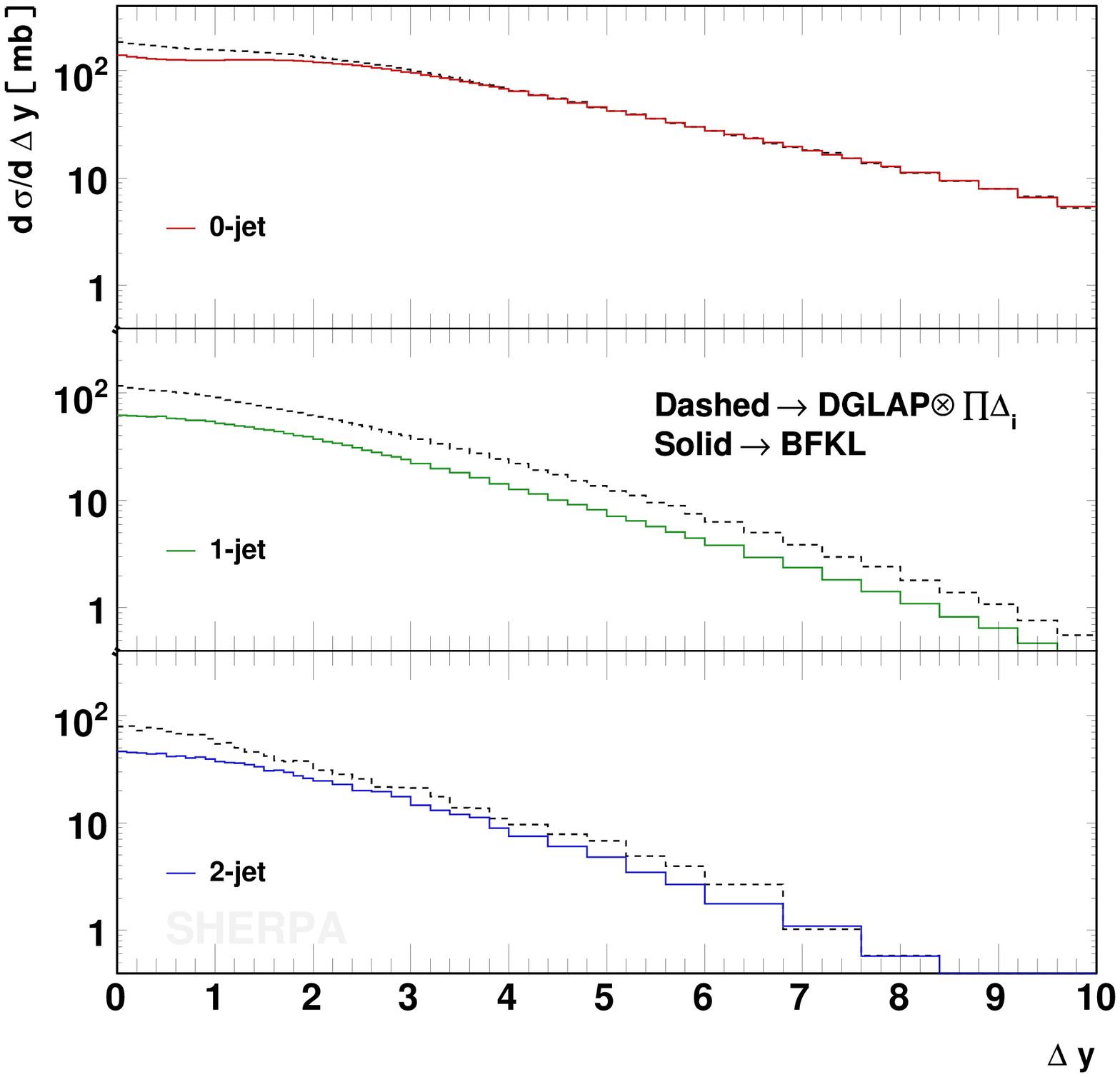}
  }{Comparison between BFKL and reweighted DGLAP matrix elements for
    the $\Delta y$-distributions.\label{dglap_bfkl_dy}}

As a next step, jet-production is investigated, comparing the results
of the new algorithm to those obtained in collinear factorisation with
on-shell matrix elements, which in the following will be denoted by DGLAP.  
The DGLAP results have been subject to the following corrections 
and constraints:
\begin{itemize}
\item ordering of final state momenta in rapidity,
\item setting $\mu_F^{(1)\,2}={\rm k}_{1\perp}^2$ 
  and $\mu_F^{(2)\,2}={\rm k}_{n\perp}^2$,
\item evaluating the coupling weight as 
  $\prod_i\alpha_s\rbr{{\rm k}_{i\perp}^2}$.
\end{itemize}
\myfigure{t}{
  \includegraphics[width=\textwidth/2-5mm]{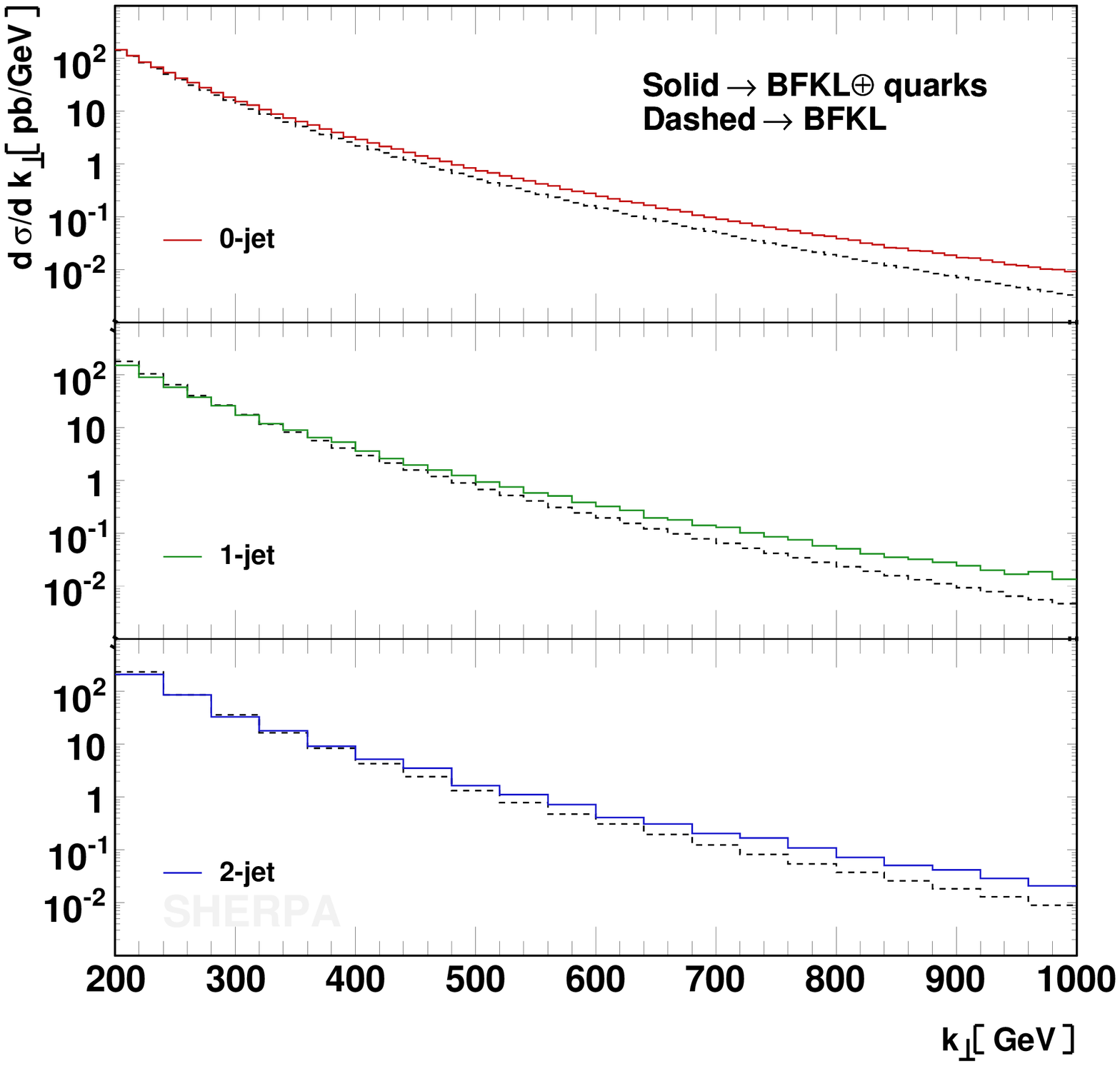}
  }{Comparison of ${\rm k}_\perp$-distributions between BFKL results 
  with and without the inclusion of quarks in the simulation.
  \label{dglap_bfkl_lkt}}
\myfigure{b!}{
  \includegraphics[width=\textwidth/2-1cm]{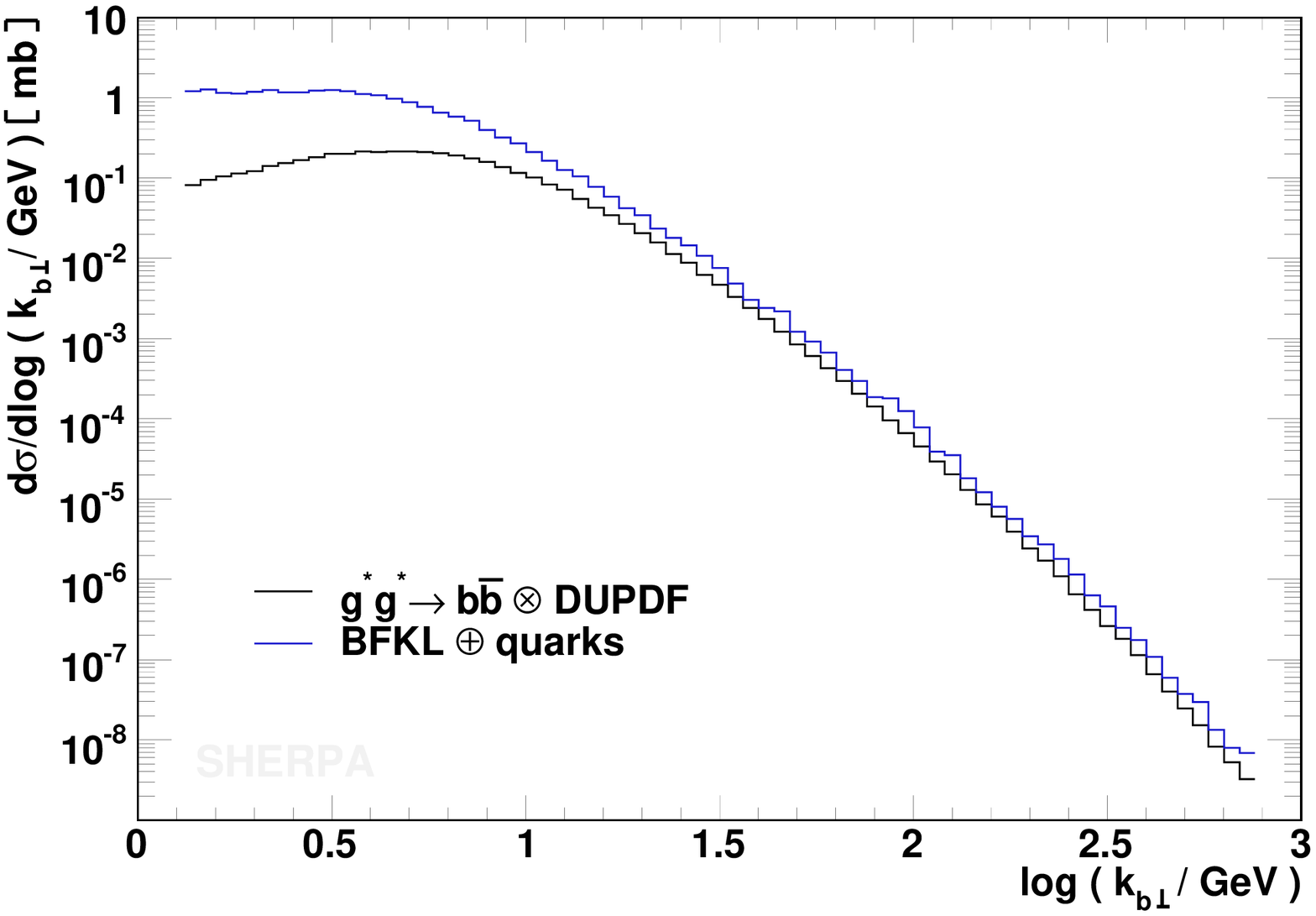}
  }{Comparison of $\log\rbr{{\rm k}_{b\perp}}$-spectra, 
  calculated either using the matrix element given in \cite{Catani:1990eg} 
  convoluted with DUPDFs or employing Eqs.~\eqref{z_kperp_factorisation_lowx}
  and~\eqref{massive_quark_splitting}.\label{hfp_jet_pt}}

However, without any $t$-channel reggeisation factor in the DGLAP 
matrix elements there are still large differences.  
Applying a $t$-channel reggeisation weight to the DGLAP calculation
results in much smaller discrepancies. The corresponding comparison for the
$\log\rbr{{\rm k}_\perp}$- and $\Delta y$-spectra is shown in
Figs.~\ref{dglap_bfkl_ikt} and~\ref{dglap_bfkl_dy}. Due to the formal
equivalence of Eqs.~\eqref{sigma_dglap_twojet} and
\eqref{z_kperp_factorisation_lowxgg} at leading logarithmic accuracy,
agreement is to be expected and can be interpreted as another
indication for the validity of the approach.
Sizable deviations occur for ${\rm k}_\perp>$ 5~GeV, which is due to
the fact that the BFKL approach is bound to describe large energy
partons only incompletely.  In order to verify this, we have reweighted 
the BFKL matrix elements with the exact matrix element obtained 
in collinear factorisation. The corresponding correction weight 
for a $2\to n$ gluonic process reads
\begin{equation*} 
  \omega=\frac{8\,n!\,M_{gg\to ng}(1,\ldots,n)}
  {\rbr{4\pi\alpha_s}^2 P_{gg}(z^{(1)})P_{gg}(z^{(2)})
  \prod_{i=2}^{n-1}\displaystyle 16\pi^2\bar{\alpha}_s/{\rm
  k}_{i\perp}^2}\;,
\end{equation*}
where the factor $n!$ occurs due to the rapidity ordering in the BFKL
approach and cancels the symmetrisation of the full DGLAP matrix
element $M_{gg\to ng}$. Performing this reweighting yields exact agreement
between the two approaches.
\myfigure{t}{
  \includegraphics[width=\textwidth/2-5mm]{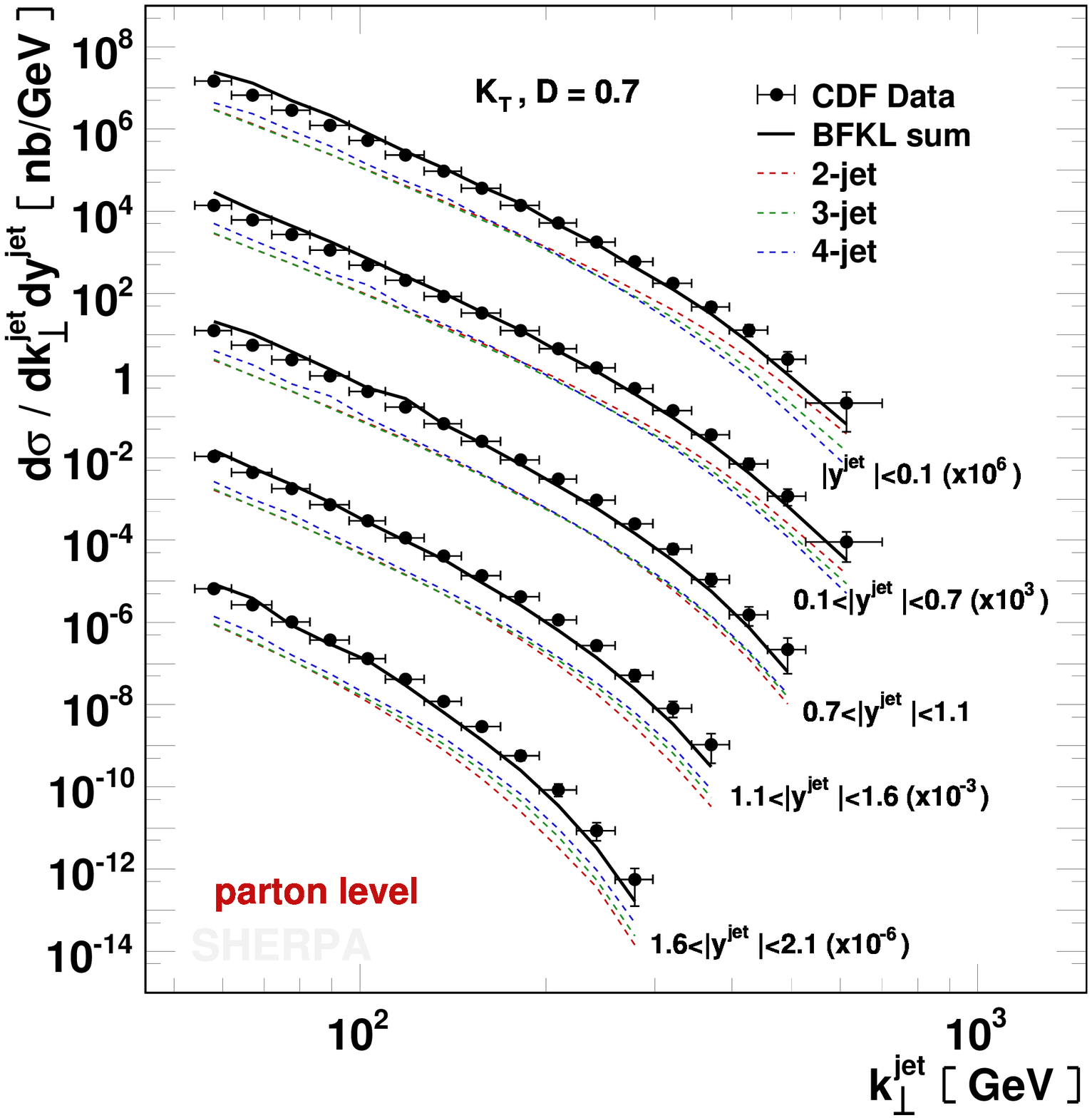}
  }{Comparison of jet-${\rm k}_\perp$-spectra with CDF data.  Details
    of the analysis can be found in \cite{Abulencia:2007ez}. Dashed
    lines show contributions from subsamples of 2- to 4-particle final
    states.\label{cdf_jet_pt}}
\myfigure{t}{
  \includegraphics[width=\textwidth/2-5mm]{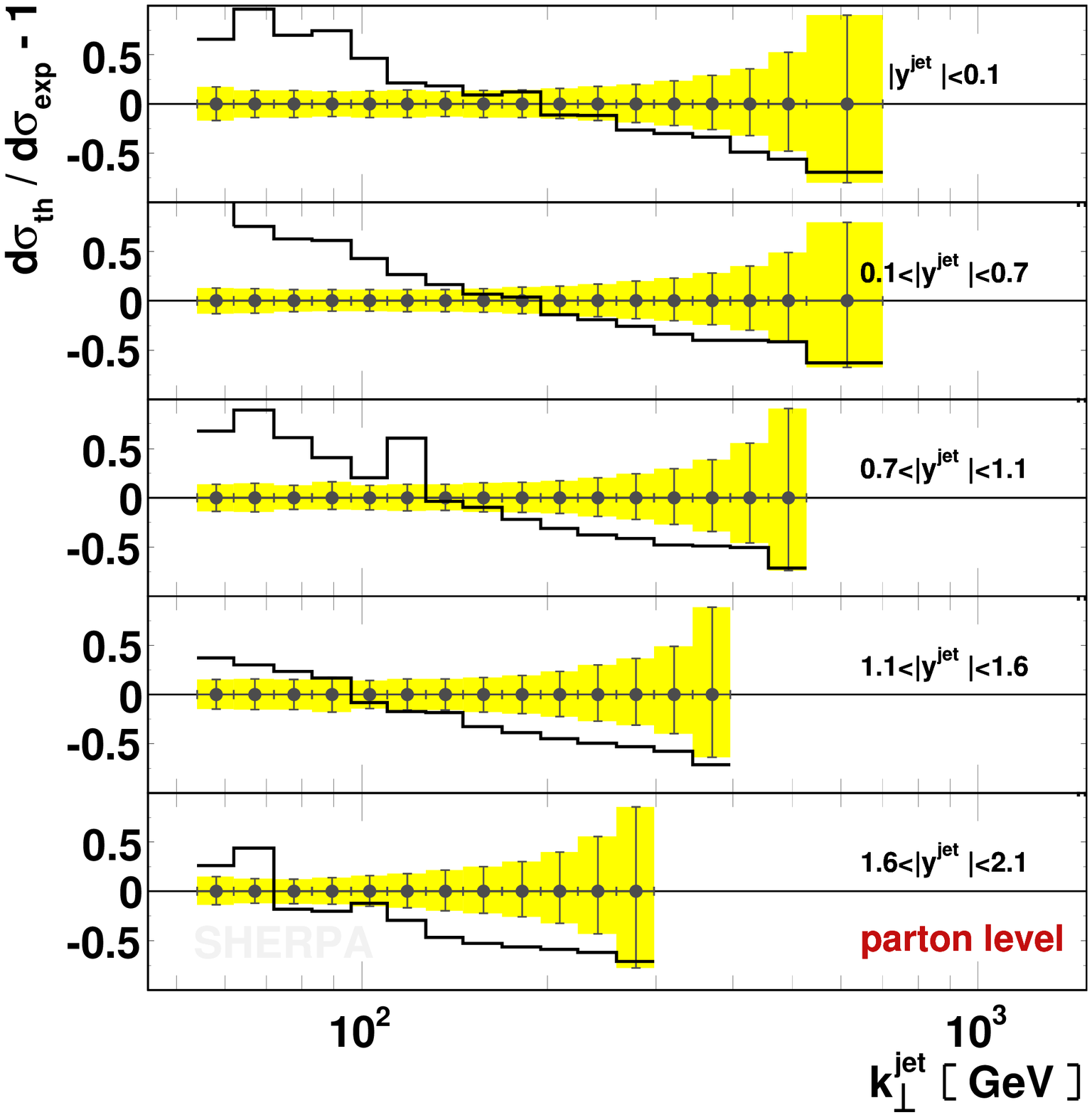}
}{Relative differences in jet-${\rm k}_\perp$-spectra compared between the
  Monte Carlo results and CDF data in Fig.~\ref{cdf_jet_pt}.
  \label{cdf_jet_ptd}}

In a next step, all possible parton splittings in the DUPDFs as well
as in the BFKL kernel have been enabled, i.e.\ 
Eq.~\eqref{z_kperp_factorisation_lowx} has been employed. As can be seen
in Fig.~\ref{dglap_bfkl_lkt}, this results in a significant change of the 
${\rm k}_\perp$-spectra of the partons in the high-${\rm k}_\perp$ region, 
which is mainly due to the fact that quarks from the PDFs tend to have 
larger energies than the gluons.
To examine the additional effect of heavy quark masses, we have compared
our results to those obtained in high-energy factorisation along the lines
of \cite{Catani:1990eg}. For this comparison we have used the full off-shell
matrix element convoluted with DUPDFs. Figure~\ref{hfp_jet_pt} shows
the $\log\rbr{{\rm k}_{\perp}}$-spectra of the heavy quarks in 
$b\bar{b}$-production. The coupling weight in the matrix element of the 
high-energy factorisation approach has been set to 
$\alpha_s\rbr{{\rm k}_{b\perp}^2}\alpha_s\rbr{{\rm k}_{\bar{b}\perp}^2}$ 
in order to match the coupling weight in our approach. We obtain
reasonable agreement with our calculation for ${\rm k}_\perp>2m_b$,
where mass effects beyond Eq.~\eqref{massive_quark_splitting} are
expected to have less impact on the results.

Finally we have compared our results to recent experimental data.
Firstly we show a comparison to data obtained by the CDF 
collaboration~\cite{Abulencia:2007ez}.  The corresponding prediction 
of jet-${\rm k}_\perp$-spectra from our MC implementation is 
shown in Fig.~\ref{cdf_jet_pt}. It fits the data considerably well, 
both in their shape and their normalisation.
Note that no $K$-factor has been employed in the calculations.
Although we observe a tilt of the distribution, which potentially arises
from missing $s$-channel contributions to quark production, this is
a quite remarkable result considering the fact that we employ a modified
LO BFKL kernel for event generation. As can be seen in Fig.~\ref{cdf_jet_ptd},
deviations are up to $\approx$50\%, which is well within the expected 
leading logarithmic accuracy. 

Secondly we compare to the decorrelation observable investigated in 
Ref.~\cite{Abachi:1996et}. As can be seen in Fig.~\ref{do_jet_decorr} 
our approach does not completely describe the data. However, the 
deviations are of similar size as in Ref.~\cite{Orr:1997im}. We stress 
that the data have not been corrected to the parton level and therefore
correlated and systematic errors might have an impact.
\myfigure{t}{
  \includegraphics[width=\textwidth/2-5mm]{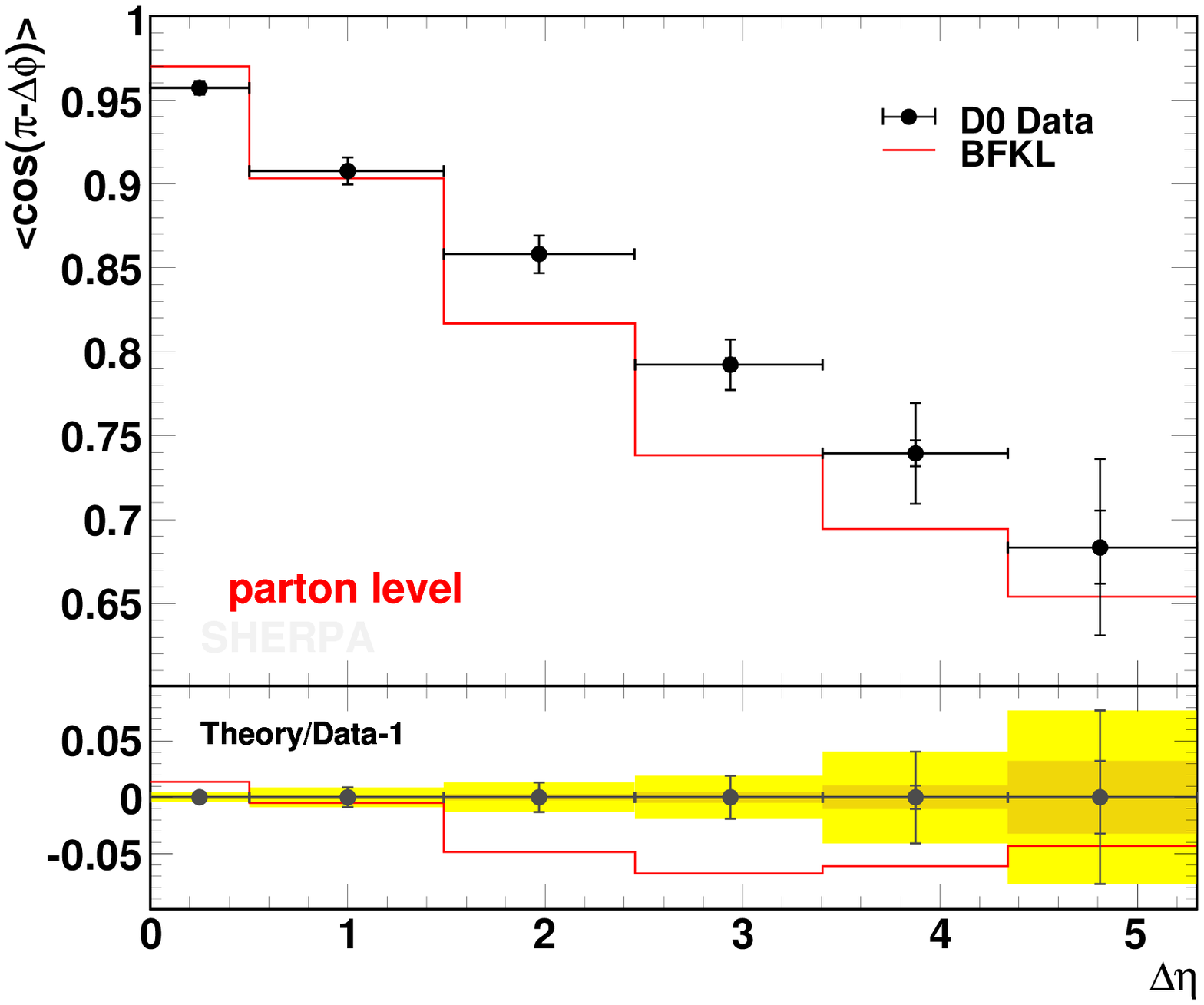}
}{Comparison of the jet decorrelation observable presented 
  in~\cite{Abachi:1996et} with D0 data. The full error bars include 
  both statistical and systematic errors, whereas statistical errors
  are independently highlighted by the smaller error bars.
  \label{do_jet_decorr}}


\mysection{Conclusions}
\label{conclusions}
In this publication we have presented a new Monte Carlo algorithm for
the description of particle production through the BFKL evolution
equation.  This has been achieved in a Markovian approach, iterating
independent emissions in order to obtain the full BFKL radiation
picture.  It has been discussed how doubly unintegrated PDFs, obtained
from conventional PDFs through the KMRW procedure can be employed
as impact factors, retaining essential features of small-$x$ physics
encoded in the BFKL equation.  In our opinion, this constitutes an
important step towards a more unified event description, which allows
to employ conventional PDFs deduced from global fits, rather than
specialised parton distributions.

The implementation of this algorithm within the framework of a
multi-purpose event generator has begun, and first results have been
discussed.  They indicate that the proposed algorithm correctly
reproduces the BFKL features visible in analytical calculations as
well as in other MC approaches.  The results also show the important
effect of a running of the coupling and of kinematical
constraints, which go beyond the LO BFKL approach.  
The realisation of the Markovian algorithm is
comparably straightforward.  Using DUPDFs obtained from collinear PDFs 
allows to compare our results for jet production to those obtained 
in the collinear factorisation approach. We found that we can obtain 
good agreement between both approaches, even for multi-parton production,
when effects that are not present in both approaches, such as
$t$-channel reggeisation and rapidity ordering, are taken into account.  
In the same framework a model for quark production, 
which is beyond the LL approximation, has been 
included and its effect on jet production has been studied. Finally,
we found that the new approach is capable to describe the production of
high-${\rm k}_\perp$ jets at the Tevatron.

This work is a first step towards a unified description of particle
production in the regime of high and low transverse momenta, i.e.\ of
jet- and minijet-production. The formalism presented here can be
extended to the simulation of multiple parton interactions, which
constitute an important part of the underlying event. Also diffractive
processes and quarkonia production may be included in the description.

\myssection{Acknowledgements}
We like to thank A.~D.~Martin and M.~G.~Ryskin for fruitful discussions.
We are especially grateful to J.~R.~Andersen for discussions concerning
the treatment of running $\alpha_s$ effects and his comments on the manuscript.
SH thanks the HEPTOOLS Marie Curie RTN (contract number MRTN-CT-2006-035505) 
for an Early Stage Researcher position.
TT thanks STFC (formerly PPARC) for an Advanced Fellowship.
Financial support by MCNet (contract number MRTN-CT-2006-035606) and BMBF
are acknowledged.

\appendix
\mysection{Alternative algorithm for phase space integration}
\label{appendix}
We explain in this section a method to integrate over the $n$-particle 
phase space, which was employed to cross-check the algorithm presented 
in Sec.~\ref{mc_procedure}. We use an iterative approach to generate 
the event topology for the process $p_a p_b\to p_1\ldots p_n$. 
For each step in the iteration we consider a $2\to 2$-scattering. 
Previous steps are taken into account by combining the particle momenta 
$p_a, p_1\ldots p_i$ into $p_{a_i}$ and thereby considering the 
$2\to 2$-process $p_{a_i} p_b\to p_i p_n$.
When denoting by $s_i=m_i^2$ and $s_{i\perp}$ the squared mass and 
squared transverse mass of the particle $i$, in the centre of mass frame 
of $p_{a_i b}$ we obtain the integration boundaries
\begin{equation*}
  \begin{split}
    E_i^{\rm max}&=\frac{1}{2\,m_{a_i b}}\rbr{s_{a_i b}+s_i-s_n}\;,\\
    {\rm k}_{i\perp}^{2\,\rm max}&=\frac{1}{4\,s_{a_i b}}
      \lambda^2\rbr{s_{a_i b},s_i,s_n}\;,
  \end{split}
\end{equation*}
where $\lambda^2\rbr{s,s_1,s_2}=\rbr{s-s_1-s_2}^2-4s_1s_2$.
The corresponding rapidity interval is fixed by
\begin{equation*}
  \begin{split}
    y_i^{max}&=\frac{1}{2}\ln\frac{1+\sqrt{1-s_{i\perp}/E_i^{\rm max\,2}}}
      {1-\sqrt{1-s_{i\perp}/E_i^{\rm max\,2}}}\;,\\
  \end{split}
\end{equation*}
and may be computed once ${\rm k}_{i\perp}^2$ is selected.
The ${\rm k}_{i\perp}^2$ selection is performed employing a divergence-free
distribution, such as $({\rm k}_{i\perp}^2)^\alpha$, where $\alpha>-1$.
Since the above boundaries are unambiguously determined, 
the $n$-particle phase space may be completely filled.

\bibliographystyle{amsunsrt_mod}  
\bibliography{bibliography}
\end{fmffile}
\end{document}